\documentclass[twocolumn,superscriptaddress,amsmath,amssymb,aps,10pt,prb,floatfix]{revtex4-2}

\usepackage{graphicx}       
\usepackage{bm}             
\usepackage{hyperref}       
\usepackage{xcolor}         

\newcommand{\ket}[1]{\left\lvert #1 \right\rangle}

\hypersetup{
    colorlinks = true,
    linkcolor = blue,
    citecolor = blue,
    urlcolor  = blue,
}

\begin{document}

\title{Detecting crossed Andreev reflection in a quantum Hall interferometer with a superconducting beam splitter}

\author{Maxime Jamotte}
\email{maxime.jamotte@uni.lu}
\affiliation{Department of Physics and Materials Science, University of Luxembourg, 1511 Luxembourg, Luxembourg}

\author{Tom Menei}
\email{tom.menei@uni.lu}
\affiliation{Department of Physics and Materials Science, University of Luxembourg, 1511 Luxembourg, Luxembourg}

\author{Manohar Kumar}
\affiliation{
 Department of Applied Physics, School of Science, Aalto University, Espoo, 02100, Finland
}

\author{Alexander Zyuzin}
\affiliation{Helsinki, Finland}

\author{Thomas L. Schmidt}
\email{thomas.schmidt@uni.lu}
\affiliation{Department of Physics and Materials Science, University of Luxembourg, 1511 Luxembourg, Luxembourg}

\date{\today}

\begin{abstract}
We study time-domain electron interferometry in a Hong-Ou-Mandel (HOM) geometry, where a thin superconductor between two quantum Hall systems acts as the beam splitter. By comparing the measurable current cross correlations at the interferometer outputs with those of a normal-conducting electronic HOM setup, we show that Andreev processes strongly affect the HOM dip. Using a combination of scattering theory and numerical tight-binding simulations for a graphene quantum Hall bar, we show that the change of charge cross correlations can be used to experimentally detect and characterize local and crossed Andreev processes. 
\end{abstract}

\maketitle


\section{Introduction}

Quantum Hall bars have become an important experimental platform to replicate the main components of optical interferometers in quantum systems obeying other statistics than the bosonic statistics of photons \cite{HeiblumFeldman,Carrega2021Anyons}. The edge states of integer and fractional quantum Hall systems can be used as chiral wave guides and gate-tunable quantum point contacts (QPCs) can act as beam splitters. Using these building blocks, many interferometer geometries, e.g., Hong-Ou-Mandel (HOM) \cite{Jonckheere2012,Rosenow,ruelle2025,Lee2022,IdrisovSchmidt}, Fabry-P\'erot \cite{two_point_contact_interferometer,nakamura2023fabryperot,Werkmeister2025,samuelsonSlowQuasiparticleDynamics2026} and Mach-Zehnder \cite{law2006,ji2003,batra2025,ghoshAnyonicBraidingChiral2024} geometries have already been investigated experimentally, which has made it possible, e.g., to measure the anyonic exchange phases of fractional quantum Hall quasiparticles. 

The exchange phases of fermions and quasiparticles are detected most conveniently in the limit of dilute beams, which can be created using tailored voltage pulses applied to the source contacts. For instance, a Lorentzian voltage pulse can be used to create ``leviton'' states, which consist of a single particle at a relatively sharp energy above a filled Fermi sea \cite{keeling08,ronettiLevitonsCorrelatedNanoscale2024}. In two-particle interferometers such as HOM or Hanbury Brown-Twiss interferometers, this makes it possible to study time-domain interferometry \cite{Lee2023,Bartolomei2022,ruelle2025}. For integer quantum Hall systems, the most basic time-domain interference signature is fermion antibunching due to the Pauli exclusion principle if two electrons are injected at the same time.

In recent years, it has become possible to couple superconductors (SCs) to quantum Hall (QH) edge states \cite{amet16,lee2017,Guel2022,vignaudEvidenceChiralSupercurrent2023,zhaoNonlocalTransportMeasurements2024}. This line of research represents the first step on the road towards non-Abelian particles, e.g., Majorana bound states or parafermion bound states, needed for topological quantum computation \cite{nayak2008topoanyons,Kitaev2003}. Andreev reflection between a superconductor and a quantum Hall state is the key physical effect behind such bound states and determines many of the experimentally measurable quantities of SC/QH interfaces \cite{MaZyuzin1993, Hoppe2000, Ostaay2011, Manesco2022,michelsen2023,kurilovich2023,kurilovich2023a,burset2025a,yakaboyluTopologicallyChargedVortices2025,bollmann2025}. To date, experimental realizations of coupled SC/QH systems are often based on measuring a ``negative downstream resistance'' in a four-terminal geometry, which results from the conversion of electrons into holes along the interface \cite{lee2017,zhaoNonlocalTransportMeasurements2024}.

In this paper, we will show that incorporating superconducting elements into quantum Hall interferometers can greatly advance the characterization of Andreev processes. The system we study is based on an electronic Hong-Ou-Mandel interferometer, in which the QPC, which usually causes transmission and reflection of electrons, is replaced by a thin superconducting wire crossing the quantum Hall bar. If the width of the superconductor is on the order of the superconducting coherence length, such a circuit element gives rise to both local and crossed Andreev reflection, in addition to normal transmission and reflection processes. We show that performing time-domain interferometry strongly changes the fermionic anti-bunching signature and will permit a direct identification of Andreev processes.

\begin{figure}
    \centering
    \includegraphics[width=0.3\textwidth]{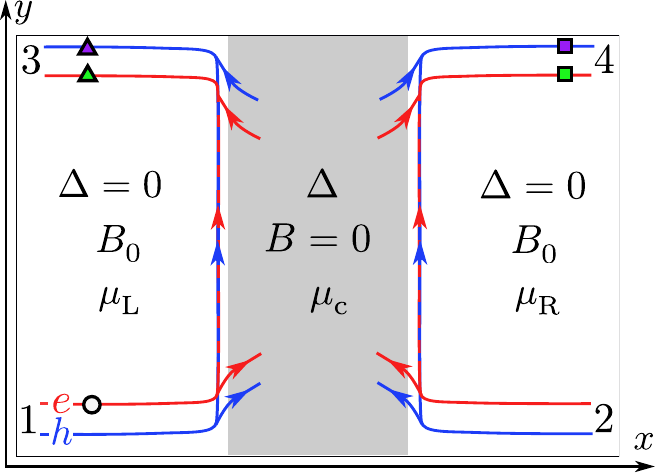}
    \caption{Sketch of the Hong-Ou-Mandel interferometer for particle- and hole-like chiral edge states, where the scattering region in the center separates two graphene sheets in a perpendicular magnetic field $B_0$. Terminals 1 and 2 are source contacts, 3 and 4 are drain contacts. The parameter $\Delta$ vanishes outside the scattering region, depicted in gray, and corresponds to either the insulator gap $\Delta_\text{I}^{}$ or the superconducting gap $\Delta_\text{S}^{}$, depending on the type of interferometer. Back gates can be used to vary the Fermi energies $\mu_{L,c,R}$ of the three regions. The indicated edge state chiralities correspond to those of a ``p-S-n'' junction, where $\mu_L > 0> \mu_c = \mu_R$.}
    \label{fig:Sketch_system}
\end{figure}

Motivated by recent experiments based on graphene as quantum Hall system, we consider a setup built on a graphene honeycomb lattice where the scattering region is composed of two quantum Hall regions separated by either a tunneling junction (acting as a normal electronic beam splitter) or a superconducting wire (superconducting beam splitter). We numerically calculate the scattering matrix of chiral states propagating along the edge of the honeycomb lattice, as sketched in Fig.~\ref{fig:Sketch_system}. While recent theory has characterized the single-particle coherence of Andreev-converted charge pulses in QH-SC-QH geometries \cite{burset2025a}, the present work addresses the two-particle sector and shows that the charge covariance in a HOM geometry provides a qualitatively new and robust signature of crossed Andreev processes that is inaccessible at the single-particle level.

We calculate time-domain interferometry signatures using Gaussian wavepackets in the source contacts as incoming states. Using a general expression for the charge noise correlator in the Bogoliubov-de Gennes (BdG) formalism to include Andreev processes, we derive charge covariances for electrons and holes arriving at different drain contacts, depending on the parameters of the beam splitter.

\section{Charge correlator}

We consider the electronic interferometer shown in Fig.~\ref{fig:Sketch_system}. As the incoming states in the source contacts 1 and 2, we consider identical wavepackets shifted by the time delay $\tau$. Using the scattering matrix $S$ of the central region, which can either be a normal insulator or a superconductor, we derive the charge correlator $\langle \hat Q_3^{} \hat Q_4^{} \rangle (\tau)$ at the drain contacts 3 and 4. The charge operator at drain contact $i = 3,4$ is defined as
\begin{equation}
    \hat Q_i^{} = e(\hat N_{ie}^{} - \hat N_{ih}^{}) = e\int d\varepsilon(\hat b_{ie}^\dagger \hat b_{ie}^{} - \hat b_{ih}^\dagger \hat b_{ih}^{}),
\end{equation}
with an electron charge $e$ and $\hat b_{i, e/h}(\varepsilon)$ being an annihilation operator of an electron or a hole (denoted by the subscripts $e/h$) with energy $\varepsilon$ at the drain contact $i$. The average in $\langle \hat Q_3^{} \hat Q_4^{} \rangle$ is taken with respect to an incoming two-particle state initialized in the leads 1 and 2 connected to the left and right edges of the scattering region.
The leads are assumed to be infinitely long, so the states are prepared at $t \to -\infty$ and reach the drain contacts at $t \to \infty$, far from the scattering region. The expression of the initial state reads
\begin{equation}\label{incoming_ee}
\begin{split}
\ket{\Psi(\tau)}_{\alpha \beta}^{}
= &\int d\varepsilon_1^{} d\varepsilon_2^{} \,
\phi(\varepsilon_1^{}) \phi(\varepsilon_2^{}) \text e^{-\text i \varepsilon_2^{} \tau}
{\hat a}_{1\alpha}^\dagger(\varepsilon_1^{}) {\hat a}_{2\beta}^\dagger(\varepsilon_2^{})
\ket{0},
\end{split}
\end{equation}
where each particle forms a wavepacket determined by the distribution $\phi(\varepsilon)$, normalized such that $\int d\varepsilon |\phi(\varepsilon)|^2 = 1$, and where $\ket 0$ is the Fermi sea. The operators ${\hat a}^\dag_{i \alpha}$ create particles in the source contacts and the indices $\alpha, \beta \in \{e, h\}$ distinguish between electrons and holes. The scattering matrix $S$ relates the creation and annihilation operators of the incoming ($\hat a$) and outgoing particles ($\hat b$) as
\begin{equation} \label{scattering}
        \begin{pmatrix}
        \hat b_{3e}^{}\\
        \hat b_{3h}^{}\\
        \hat b_{4e}^{}\\
        \hat b_{4h}^{}
    \end{pmatrix}
    =
    \begin{pmatrix}
        r_{31}^{ee} & r_{31}^{eh} & t_{32}^{ee} & t_{32}^{eh}\\
        r_{31}^{he} & r_{31}^{hh} & t_{32}^{he} & t_{32}^{hh}\\
        t_{41}^{ee} & t_{41}^{eh} & r_{42}^{ee} & r_{42}^{eh}\\
        t_{41}^{he} & t_{41}^{hh} & r_{42}^{he} & r_{42}^{hh}\\
    \end{pmatrix}
    \begin{pmatrix}
        \hat a_{1e}^{}\\
        \hat a_{1h}^{}\\
        \hat a_{2e}^{}\\
        \hat a_{2h}^{}
    \end{pmatrix}.
\end{equation}
This yields the following expression for the charge correlator,
\begin{align}\label{Q3Q4_general}
    &\langle \hat Q_3^{} \hat Q_4^{} \rangle_{\alpha \beta}^{} (\tau) = \text{Re} \sum_{\gamma,\delta}q_\gamma^{} q_\delta \Big[ \\
    &\phantom{+}
    \int d\varepsilon |t_{32}^{\gamma \beta}(\varepsilon)|^2 |\phi(\varepsilon)|^2 \int d\varepsilon' |t_{41}^{\delta \beta}(\varepsilon')|^2 |\phi(\varepsilon')|^2 \notag \\
    &+ \int d\varepsilon |r_{31}^{\gamma \alpha}(\varepsilon)|^2   |\phi(\varepsilon)|^2 \int d\varepsilon' |r_{42}^{\delta \alpha}(\varepsilon')|^2 |\phi(\varepsilon')|^2 \notag \\
    &-2 \int d\varepsilon  |{\phi(\varepsilon)}|^2 (r^{\gamma\alpha}_{31}(\varepsilon))^* t^{\gamma\beta}_{32}(\varepsilon) \text e^{\text i\varepsilon\tau}\  \times \notag \\
    &\quad \ \ \int d\varepsilon' |{\phi(\varepsilon')}|^2 \left(r^{\delta\beta}_{42}(\varepsilon') \right)^*  t^{\delta\alpha}_{41}(\varepsilon')\ \text e^{-\text i\varepsilon'\tau}\Big], \notag
\end{align}
where $\alpha,\beta,\gamma,\delta \in \{e,h\}$, $q_\gamma^{}, q_\delta^{}$ are the charges of the particles at the drain contacts 3 and 4, respectively, and $q_e^{} = -e$, $q_h^{} =e$. The first two terms in Eq.~\eqref{Q3Q4_general} describe classical contributions associated with independent scattering events and are therefore insensitive to the time delay $\tau$. The third term arises from two-particle interference between the indistinguishable particles and carries the $\tau$-dependence via the phase factors $e^{\pm \text i \varepsilon \tau}$. This interference term is responsible for the HOM–type correlations. In the following, we will calculate and discuss the charge correlator in the cases when the beam splitter is an insulator or a superconductor.

\section{Models}

In this section, we define the Hamiltonians of the different regions that are combined as building blocks to form the interferometers. This Hamiltonian will then serve as a basis for a numerical calculation of the scattering matrix $S$. 

\subsection{Graphene in the quantum Hall regime (QHG)}

Near half-filling, the band structure of the $\pi$-bands of graphene can be approximated by a tight-binding Hamiltonian on a honeycomb lattice. The Hamiltonian of graphene in the quantum Hall regime is given by
\begin{equation}\label{H_QHG}
\begin{split}
    \hat H_\text{QHG}^{} =  - J \sum_{\mathbf r,j} \text{e}^{-\text i \mathbf A \cdot \boldsymbol{\delta}_j^{}} \hat A_{\mathbf r}^\dagger \hat B^{}_{\mathbf r_j^{}}+ \text{H.c.},
\end{split}
\end{equation}
where $\mathbf A(\mathbf r)$ is the vector potential and the vectors connecting nearest neighbors are denoted by $\boldsymbol{\delta}_1^{} = (0,-a)$, $\boldsymbol{\delta}_2^{} = (\sqrt 3 a/2,a/2)$, and $\boldsymbol{\delta}_2^{} = (-\sqrt 3a/2,a/2)$ with the lattice spacing $a$. We used the notation $\mathbf r = (x,y)$ and $\mathbf r_j^{} = \mathbf r + \boldsymbol{\delta}_j^{}$. The operators $\hat A_{\mathbf{r}}$ and $\hat B_{\mathbf{r}}$ annihilate electrons on the $A$ and $B$ sublattices, respectively. We neglect the Zeeman effect because it is small compared to the orbital effect at the magnetic field strengths in the quantum Hall regime. Moreover, we neglect spin-orbit coupling because it is weak in graphene. These assumptions allow us to consider spin-degenerate models. 

\begin{figure}
    \centering
    \includegraphics[width=\linewidth]{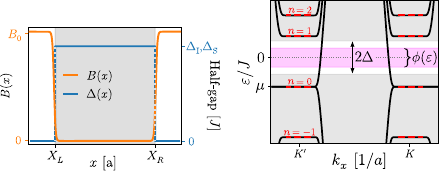}
    \caption{\emph{(a)} Position dependence of the magnetic field (orange) and the energy gap $\Delta_\text{I}^{}$ or $\Delta_S^{}$ (blue). \emph{(b)} Lead spectrum displaying relativistic Landau levels (red lines) at energies $\varepsilon_n^{\text{LL}} = \pm v_\text{F}^{} \sqrt{2 \hbar|B n|} + \mu$, where $\mu$ is the chemical potential in the leads. The white area represents the gap due to the $\Delta = \Delta_\text{I}^{}$ in Eq.~\eqref{H_ins} or $\Delta = \Delta_\text{S}^{}$ in Eq.~\eqref{H_SC}. The pink area highlights the energy window of the components of the wavepacket \eqref{incoming_ee}.}
    \label{fig:Sketch_spectra}
\end{figure}

A gapped scattering region separates two QHG regions and acts as a beam splitter, allowing for forward and back-scattering. Back gates can be used to shift the Fermi levels on the two QHG regions independently to the values $\mu_{L,R}$. Since we are mainly interested in using a superconductor as a beam splitter, we assume that the magnetic field is screened in the scattering region, as shown in Fig.~\ref{fig:Sketch_spectra}a. This leads to the phenomenological magnetic field profile,
\begin{align}\label{eq:B}
    B(x) = \frac{B_0^{}}{2} \left[2-\tanh\left(\frac{x-X_L^{}}{d}\right) + \tanh\left(\frac{x-X_R^{}}{d}\right)\right].
\end{align}
The parameter $d$ is the magnetic penetration depth and controls the steepness of the magnetic field at the interfaces. $X_L^{}$ and $X_R^{}$ are the positions of the left and right interfaces, respectively. The magnetic field approaches $B_0$ far from the beam splitter. Our choice of a hyperbolic tangent allows us to account for the Meissner effect, which screens the magnetic field exponentially in the superconductor. By choosing an adequate gauge, such as the Landau gauge $\mathbf A = (-B(x)y,0,0)$, translation invariance along the $x$-axis in the leads is preserved. A representative lead energy spectrum is shown in Fig.~\ref{fig:Sketch_spectra}b, where the Landau levels are highlighted (in red) and appear at energies $\varepsilon_n^\text{LL} = \pm v_\text{F}^{} \sqrt{2\hbar|n B_0^{}|} \ (n \in \mathbb Z)$, where $v_\text{F}^{}$ is the Fermi velocity, with respect to the chemical potential $\mu$. Note that because of spin degeneracy, the $n^\text{th}_{}$ Landau level corresponds to the filling factor $\nu = 2n$. In this panel, we have added the energy window (in pink) over which we build the two incoming wavepackets. For the rest of this work, we place ourselves in the limit of large system sizes in comparison to the magnetic length $\ell_B^{} = \sqrt{\hbar/e|B_0^{}|}$. Importantly, due to computational constraints, the numerical simulations are performed on systems with a smaller number of lattice sites than typical experimental devices. However, all relevant dimensionless parameters—such as the ratios between system size, magnetic length, and superconducting coherence length—are preserved. As a result, the simulations probe the same physical regime as in realistic setups. We note that the range of the superconducting gap is chosen such that the BCS coherence length $\xi_0 = \hbar v_\text{F}^{} /\Delta_\text{S}^{}$ is comparable to the widths of superconducting fingers in recent experiments \cite{lee2017}.

\subsection{Insulator as the beam splitter}

We use a normal insulator between two QHG regions as a reference point for the comparison of the current cross correlations. This case corresponds to the normal HOM geometry with a quantum point contact acting as beam splitter. To avoid interface effects, we model the insulator at $x \in [X_L, X_R]$ using a honeycomb lattice with a staggered sublattice potential, i.e., onsite potentials $+\Delta_\text{I}^{}$ and $-\Delta_\text{I}^{}$ at the A and B sites, respectively. The Hamiltonian of the scattering region then reads
\begin{equation}\label{H_ins}
    \hat H_\text{I}^{} = \hat H_\text{QHG}^{} + \sum_{\mathbf{r}} \Delta_\text{I}^{}(x) (\hat A^\dagger_{\mathbf{r}} \hat A^{}_{\mathbf{r}}-\hat B^\dagger_{\mathbf{r}} \hat B^{}_{\mathbf{r}}).
\end{equation}
The spatial dependence of this onsite potential $\Delta_I$ is depicted in Fig.~\ref{fig:Sketch_spectra}a by the solid blue line (for $\Delta = \Delta_\text{I}^{}$). The staggered sublattice potential opens a topologically trivial gap of width $2\Delta_\text{I}^{}$ in the energy spectrum of the scattering region, centered around the charge neutrality point. This gap is depicted in Fig.~\ref{fig:Sketch_spectra}b by the white area centered around $\varepsilon = 0$. As described by Eq.~\eqref{eq:B} and shown in Fig.~\ref{fig:Sketch_spectra}a, we neglect for simplicity the magnetic field inside the insulator, which corresponds to assuming that $\ell_B \gg \hbar v_F/\Delta_I$.

\subsection{Superconductor as the beam splitter}

For the superconducting interferometer setup, the beam splitter is modeled as a type-II superconductor, where the strength of the magnetic field $B_0^{}$ is supposed to be below its upper critical magnetic field. We assume that its action on the charges within the SC does not qualitatively change the results. To accommodate the introduction of Bogoliubov quasiparticles, we use the Hamiltonian of the QHG and SC regions:
\begin{equation}\label{H_SC}
\begin{split}
    \hat H_\text{SC} = \hat H_\text{QHG}^{}
    +  \sum_{\mathbf r } \Delta_\text{S}^{}(x) \left(\hat c_{ \mathbf r \uparrow}^\dagger \hat c_{\mathbf r\downarrow }^\dagger + \hat c_{\mathbf r\downarrow}^{} \hat c_{\mathbf r \uparrow}^{}  \right),
\end{split}
\end{equation}
where the order parameter of the superconductor is chosen to be real, and $\hat c^{(\dagger)}_{\mathbf r} = \hat A^{(\dagger)}_{\mathbf r}$ or $\hat B^{(\dagger)}_{\mathbf r}$, depending on $\mathbf r$ being a $A$-site or $B$-site position, respectively. Since this $s$-wave pairing conserves spin, we can exploit spin degeneracy and use two-component Nambu spinors consisting of spin-$\uparrow$ electrons and spin-$\downarrow$ holes for the numerical calculation using the BdG formalism \cite{BTK1982,Akhmerov2007,Hoppe2000,Manesco2022}. The spatial distribution of the pairing strength is depicted by the blue line in Fig.~\ref{fig:Sketch_spectra}a for $\Delta = \Delta_\text{S}^{}$. It creates a gap of $2\Delta_\text{S}^{}$ in the spectrum of the superconducting beam splitter, as depicted in Fig.~\ref{fig:Sketch_spectra}b by the white area.

\section{QHG-Insulator-QHG interferometer}

We begin by demonstrating that our numerical simulations of a QHG-Ins-QHG junction reproduce known results for electronic beam-splitters. We prepare chiral quantum Hall edge states propagating along the bottom left and top right edges towards the insulating region. The insulator band gap $\Delta_I$ determines the transmission and reflection amplitudes, giving rise to different electronic interferences. The incoming states are built from identical wavepackets in an energy window between the zeroth and the first Landau levels, making the two electrons indistinguishable. For each energy $\varepsilon$ in this window, we are able to compute the scattering matrix $S(\varepsilon)$, using the package \textsc{Kwant} \cite{Kwant}.

\begin{figure}
    \centering
    \includegraphics[width=\linewidth]{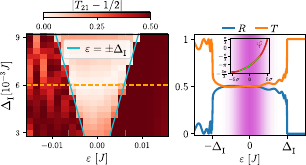}
    \caption{(\emph{a}) Map of the absolute difference of the transmission amplitude $T = |t_{ee}^{}|^2$ from $1/2$ as a function of energy $\varepsilon$ and the insulator gap $\Delta_\text{I}^{}$. The cyan lines represents the lower and upper limit of the energy gap, $\varepsilon = \pm \Delta_\text{I}^{}$. (\emph{b}) Reflection (blue) and transmission (orange) probabilities as function of energy for $\Delta_\text{I}^{} = 6 \times 10^{-3} J$ (indicated by the orange dashed line in panel a) and $L_\text{Ins}^{} = 232a$. The pink area represents the energy window of the wavepacket with the Gaussian envelope $\phi(\varepsilon)$. \emph{Inset:} Phase difference $\varphi = \theta_t^{} - \theta_r^{}$ (in red) across the part of the energy window where the $\phi(\varepsilon)$ is non-negligible ($[-5\sigma,5\sigma]$). The parabolic fit (green) over $\varepsilon \in [-3\sigma,3\sigma]$, providing $\kappa_{1,2}^{}$, is also visible.}
    \label{fig:Tampl}
\end{figure}

When one electron from contact 1 ($\hat{a}_1$) and one from contact 2 ($\hat{a}_2$) are sent toward the beam splitter, the outgoing modes $\hat{b}_3$ and $\hat{b}_4$ can be described by the scattering matrix,
\begin{align}
\begin{pmatrix}
\hat{b}_3^{} \\
\hat{b}_4^{}\\
\end{pmatrix}&=
\underbrace{
\begin{pmatrix}
r_{31}^{} & t_{32}^{} \\
t_{41}^{} & r_{42}^{}
\end{pmatrix}
}_{S}
\begin{pmatrix}
\hat{a}_1^{} \\
\hat{a}_2^{}
\end{pmatrix}.
\end{align}
From the unitarity of the scattering matrix $S$, one can deduce the following relations between the transmission and reflection amplitudes:
\begin{equation}\label{scatt_cond}
    S^\dagger_{} S = I = S S^\dagger_{}
\;\;\Rightarrow\;\;
\begin{cases}
|r_{31}^{}|^2 + |t_{41}^{}|^2 = 1, \\
|r_{42}^{}|^2 + |t_{32}^{}|^2 = 1, \\
t_{32}^{}\, r_{31}^\ast = - r_{42}^{}\, t_{41}^\ast,\\
t_{32}^{}\, r_{42}^* = - r_{31}^{} \,t^{*}_{41}.
\end{cases}
\end{equation}
From the last two equations, we deduce that $|t_{32}^{}| = |t_{41}^{}| = |t|$ and $|r_{31}^{}| = |r_{42}^{}| = |r|$. Hence, the charge correlator simplifies to
\begin{align}
\frac{\langle \hat Q_{3}^{}\hat Q_4^{} \rangle(\tau)}{e^2} 
=& \quad \left(\int d\varepsilon |t(\varepsilon)|^2 |\phi(\varepsilon)|^2 \right)^2 \\
 &+\left(\int d\varepsilon |r(\varepsilon)|^2 |\phi(\varepsilon)|^2 \right)^2\\
&+ 2\left|\int d\varepsilon\, t(\varepsilon)\, r^* (\varepsilon) \text e^{\text i \varepsilon \tau} \, |\phi(\varepsilon)|^2 \right|^2.
\end{align}
The energy-dependent reflection and transmission probabilities are shown in Fig.~\ref{fig:Tampl}b and for our chosen parameters of the beam splitter length and band gap $\Delta_\text{I}^{}$, there is a wide range of energies inside the band gap, which with both reflection and transmission probabilities are approximately equal -- the optimal scenario for Hong-Ou-Mandel interferometry. The asymmetry between $\varepsilon < 0$ and $\varepsilon > 0$ arises from the finite chemical potential $\mu = -0.061, J$, which shifts the Fermi level away from the charge neutrality point. This breaks the symmetry between states above and below the Fermi energy, leading to an asymmetric energy dependence of the transmission and reflection coefficients.

\begin{figure}
\centering 
\includegraphics[width=\linewidth]{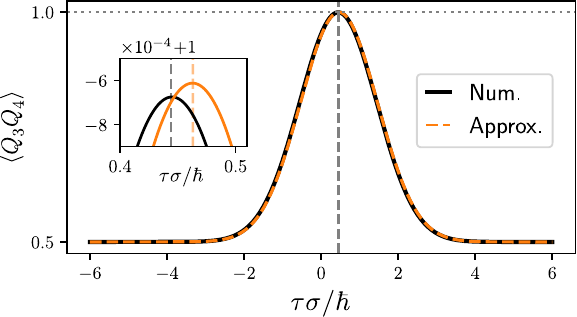}
\caption{Charge correlator $\langle \hat Q_3^{} \hat Q_4^{} \rangle$ versus the time delay $\tau$ between the two incoming electrons. The numerical results are shown by the black line, while the orange dashed line corresponds to the analytical expression given by Eq.~\eqref{Q_34_final}. The shift of the interference signal is highlighted by the vertical dashed line, at $\sigma \tau/ \hbar \approx 0.44$. The inset zooms on the top of the peak to show that does not reach 1 as expected in an ideal case. The vertical dashed lines indicate the values $\sigma \tau /\hbar$ at which the numerical and analytical approximation reach their maxima; at $\sigma \tau /\hbar  \approx 0.44$ and $0.46$, respectively.}
\label{fig:Q3Q4}
\end{figure}

The deviations from the point $t^2 = r^2 = 1/2$ away from the average energy of the wave packet $\varepsilon_0^{}$ are weighted by the tail of the wavepacket in the integral over energies. Hence, for a wavepacket narrow enough in energies (depicted by the pink area in Fig.~\ref{fig:Tampl}) they have a negligible impact on the integration over energies. This allows us to further simplify the previous equation. Indeed, if $r(\varepsilon) = \sqrt{R} \text e^{\text i \theta_r^{}(\varepsilon)}$ and $t(\varepsilon) = \sqrt{T}\text e^{\text i \theta_t^{}(\varepsilon)}$, we finally obtain
\begin{equation}\label{Q_34}
\begin{split}
\frac{\langle \hat Q_{3}^{} \hat Q_4^{} \rangle(\tau)}{e^2} = R^2+T^2+ 2RT\left|\int d\varepsilon\,
|\phi(\varepsilon)|^2\,
\text e^{\text i (\varphi(\varepsilon)-\varepsilon \tau)}
\right|^2,
\end{split}
\end{equation}
where $\varphi(\varepsilon) = \theta_t^{}(\varepsilon) - \theta_r^{}(\varepsilon)$ is the phase difference between $t_{ee}^{}$ and $r_{ee}^{*}$. A constant value of $\varphi(\varepsilon)$ drops out when taking the absolute value. Additionally, as shown in Fig.~\ref{fig:Tampl}, the dependence of $\varphi(\varepsilon)$ on energy can be well approximated by a quadratic function $\kappa_2^{} \varepsilon^2 + \kappa_1^{} \varepsilon$ within the energy integration window. Finally, assuming a Gaussian wavepacket
\begin{equation}
    \phi(\varepsilon) = \frac{1}{\sqrt[4]{\pi \sigma^2}} \text e^{-(\varepsilon-\varepsilon_0^{})^2/2\sigma^2},
\end{equation}
we find that 
\begin{equation}\label{Q_34_final}
\begin{split}
\frac{\langle \hat Q_{3}^{} \hat Q_4^{} \rangle(\tau)}{e^2} =& \frac{2RT}{\sqrt{1 + \kappa_2^2 \sigma^4_{}}} \exp\left(-\frac{\sigma^2(\kappa_1^{}-\tau)^2}{2(1 + \kappa_2^2 \sigma^4_{})}\right) \\
&+ R^2+T^2.
\end{split}
\end{equation}
The charge correlator is a shifted Gaussian function with a maximum at $\tau \approx \kappa_1^{}$. Conversely to the ideal case of constant amplitudes, the energy dependence of $\varphi(\varepsilon)$ leads to a reduction in the visibility of the HOM dip, as shown by the prefactor. This translates to a partial loss of indistinguishability between the two incoming electrons. The comparison between the numerical results and the analytical expression of the charge correlator is shown in Fig.~\ref{fig:Q3Q4}. We see that the numerical results fit the analytical approximation very nicely. The corresponding dip in the autocorrelation function of the currents was observed in electron HOM experiments \cite{Feve2013,Feve2015}.
The inset of Fig.~\ref{fig:Q3Q4} shows the same plot zoomed in around the maximum of the peak, highlighting the reduction in the visibility of the HOM peak as well as the shift away from $\tau = 0$ due to the energy dependence of $\varphi$. 

In summary, we conclude that the QHG-Ins-QHG junction provides a controlled benchmark as it reproduces the standard HOM response of a normal beam splitter \cite{Jonckheere2012,Feve2013}. In particular, we recover the expected anti-bunching signal when injecting two indistinguishable electrons, which is reflected by the maximum of the charge correlator at small time delay. Our analysis also shows that the residual energy dependence of the scattering phase broadens this peak and reduces its height, revealing a small departure from the ideal case of perfectly energy-independent amplitudes. Furthermore, the excellent agreement between the numerical results and the analytical approximation confirms that the interferometric signal remains directly related to the profile of the incoming wavepacket. This validates our numerical framework and provides a solid reference point for identifying superconducting effects in the following section.

\section{QHG-SC-QHG interferometer}

\begin{figure}
    \centering
    \includegraphics[width=\linewidth]{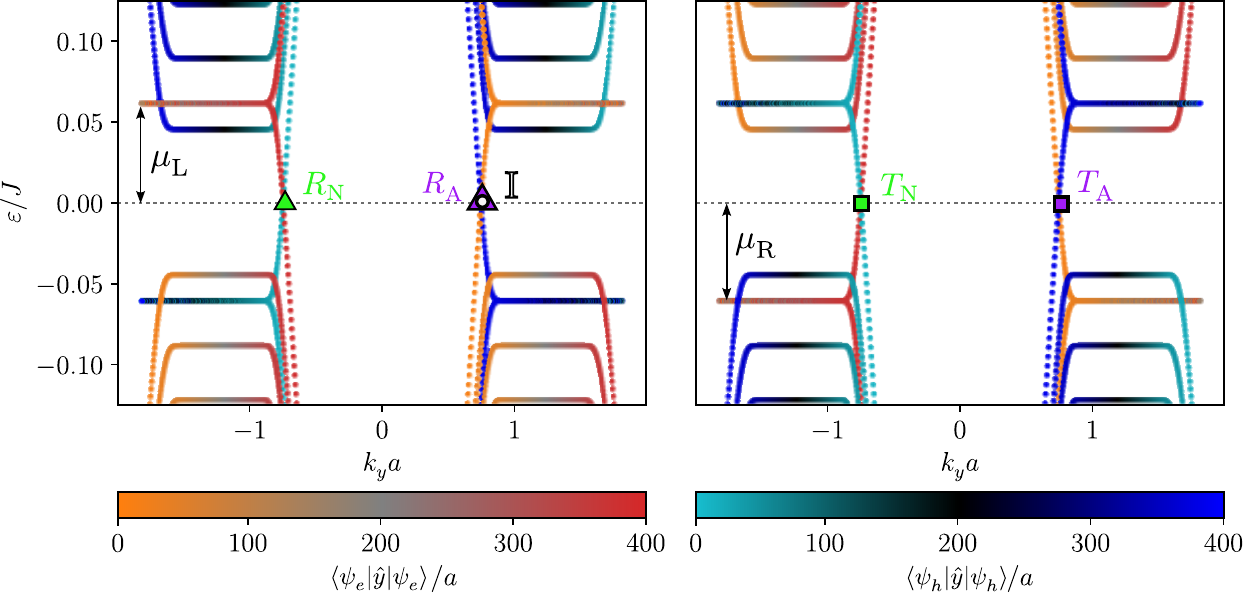}
    \caption{Spectra of the left (panel \textbf{a}) and right (panel \textbf{b}) QHG regions, where electron and hole branches are distinguished. The color scale indicates the average $y$ position of the wavefunction associated with each point $(k_y,\varepsilon)$. Electron branches are shown in orange-gray-red and hole branches in cyan-black-blue, corresponding respectively to states localized near the bottom, center, and top of the sample. The opposite chemical potentials $\mu_\text{L}$ and $\mu_\text{R}$ lead to an alignment between the electron-like branch of the left QHG region and the hole-like branch of the right QHG region. The markers correspond to those in Fig.~\ref{fig:Sketch_system}: the white circle denotes an incoming electron along the lower left edge, while the green and magenta symbols indicate the outgoing edge states associated with normal and Andreev reflection/transmission processes.}
    \label{fig:spectra_SC}
\end{figure}

In this section, we show that Andreev processes can be unambiguously identified in the HOM signal of a QHG-SC-QHG junction, and, in particular, that crossed Andreev processes can be directly detected. For this purpose, we consider a ``p-S-n'' configuration, in which the chemical potentials of the left and right QHG regions have opposite signs, $\mu_\text{L}=-\mu_\text{R}=0.061J$. As shown in Fig.~\ref{fig:spectra_SC}, this choice aligns the electron-like branch of the left QHG region (orange-red lines in panel a) with the hole-like branch of the right QHG region (cyan-blue lines in panel b), which favors crossed Andreev processes \cite{Zhang2019}. At these chemical potentials, the Fermi energy remains between the zeroth and first Landau levels in both leads, so that only the lowest edge channels participate in transport. The markers in Fig.~\ref{fig:spectra_SC} should be read together with the interferometer sketched in Fig.~\ref{fig:Sketch_system}: the white circle labeled \textrm{I} denotes an electron injected from the left source and propagating chirally toward the beam splitter along the lower edge, as sketched by the red arrow in the left region in Fig.~\ref{fig:Sketch_system}. After scattering at the superconducting region, this incoming excitation can couple to the outgoing chiral states indicated by the colored markers; in particular, the state marked by the magenta square ($T_\text{A}^{}$) in panel b illustrates that the incoming electron can be converted into a hole-like excitation in the right QHG (blue arrow in Fig.~\ref{fig:Sketch_system}), which is the signature of crossed Andreev transmission. We use the same logic for all possible processes: the green square ($T_\text{N}^{}$) corresponds to a local Andreev reflection into the electron-like branch of the right QHG, while the markers in panel a identify the reflected channels in the left QHG, namely normal reflection $R_\text{N}^{}$ within the electron-like branch and local Andreev reflection $R_\text{A}^{}$ into the hole-like branch. In this way, Fig.~\ref{fig:spectra_SC} provides a spectral interpretation of the scattering processes sketched in Fig.~\ref{fig:Sketch_system}.

\begin{figure}
    \centering
    \includegraphics[width=\linewidth]{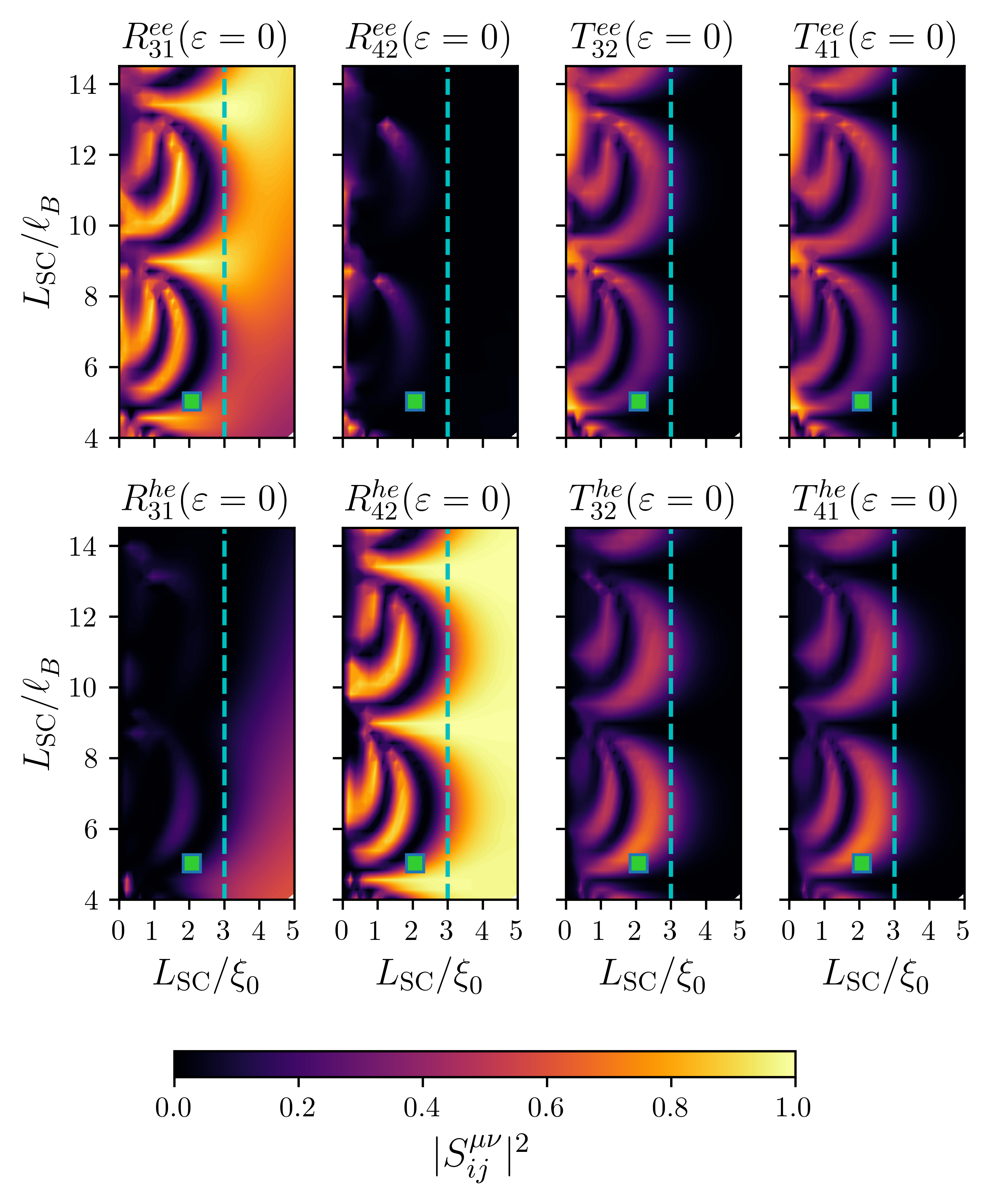}
    \caption{Scattering probabilities $|S_{ij}^{\mu e}|^2$ (for $\mu \in \{e, h\}$) as a function of the length scale ratios $L_\text{SC}^{}/\xi_0$ and $L_\text{SC}/\ell_B^{}$. The cyan line indicates $L_\text{SC}=3\xi_0$. The green squares indicate the working point selected for Fig.~\ref{fig:Scatt_energy}.}
    \label{fig:ampl_vs_Detla_LSC}
\end{figure}

To identify the impact of superconductivity on the HOM signal, we first analyze the scattering probabilities associated with the different processes. This allows us to identify a regime in which normal and Andreev processes coexist with comparable weights, which is essential for observing superconductivity-induced modifications of the HOM interference signal. Indeed, if either the normal or the Andreev channels dominate, the HOM signal is expected to remain close to that of a purely normal conducting interferometer, as both outgoing particles will be indistinguishable electrons or holes. We first determine the optimal working point in the parameter space of the superconducting beam splitter, which is spanned by the superconducting gap $\Delta_\text{S}^{}$ and the length of the superconducting region $L_\text{SC}^{}$, with fixed chemical potentials in the normal regions and a fixed magnetic field (still in the limit $a \ll \ell_B^{} \ll L_\text{QHG}^{},L_y^{}$). For $L_y^{}/\ell_B^{} = 22$ and $\mu_\text{QHG,l}^{} = 0.061\,J$, we compute the corresponding scattering probabilities $|S_{ij}^{\mu e}|^2$, shown in Fig.~\ref{fig:ampl_vs_Detla_LSC}. Two main features emerge from this parameter scan. First, once $L_\text{SC} \gtrsim 3\xi_0$, transmission processes are strongly suppressed, consistent with the exponential attenuation $\exp(-L_\text{SC}^{}/\xi_0)$ expected in a gapped region, where the associated decay length is the superconducting coherence length $\xi_0 = \hbar v_\text{F}/\Delta_\text{S}$. Moreover, the dependence on this scale ratio is consistent with the recurrent lobes visible in the scans. Second, below this critical line, normal and Andreev channels coexist with comparable weights. This is the most relevant regime for observing clear superconductivity-related effects in the HOM interference signal.

\begin{figure}
    \centering
    \includegraphics[width=\linewidth]{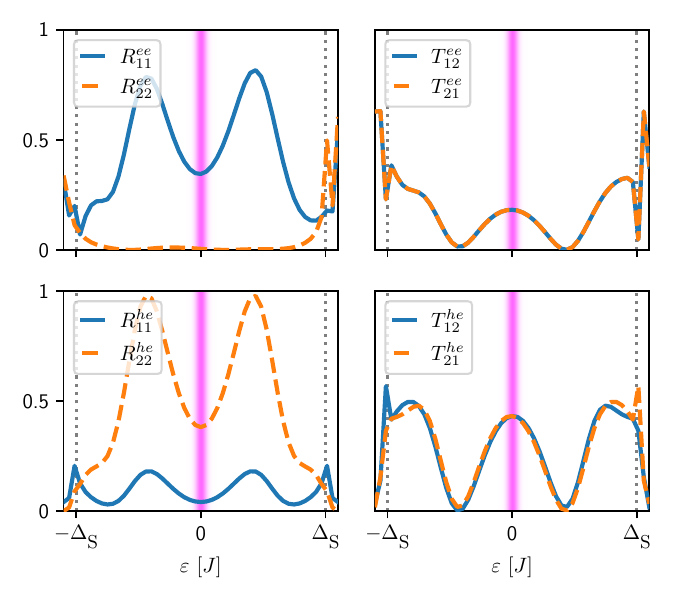}
    \caption{Scattering amplitudes $|S_{ij}^{\mu e}|^2$ as a function of the energy for the ratios $L_\text{SC}^{}/\ell_B^{} = 5$ and $L_\text{SC}^{}/\xi_0 = 2$. The magenta area indicates the Gaussian weight over the energy window that the wavepacket is defined on.}
    \label{fig:Scatt_energy}
\end{figure}

Based on Fig.~\ref{fig:ampl_vs_Detla_LSC}, we choose $L_\text{SC}^{}/\ell_B^{}=5$ and $L_\text{SC}^{}/\xi_0 = 2$ as suitable working points, indicated by the green squares in all panels. This choice of parameters aligns with recent experimental efforts using narrow SC fingers with widths of a few coherence lengths \cite{lee2017,Guel2022}. The corresponding energy dependence of the scattering probabilities is shown in Fig.~\ref{fig:Scatt_energy}. Over the energy window selected by the Gaussian envelope of the incoming wavepacket, all relevant probabilities remain finite and vary only moderately, which allows normal and Andreev processes to contribute on a comparable footing to the charge correlator over the energy window.

Using these amplitudes, we compute the covariance 
$\langle \delta \hat Q_3^{} \delta \hat Q_4^{} \rangle_{ee}^{} = \langle \hat Q_3^{} \hat Q_4^{} \rangle_{ee}^{} - \langle \hat Q_3^{} \rangle_{ee}^{} \langle \hat Q_4^{} \rangle_{ee}^{}$ between the charge operators $\hat Q_3^{}$ and $\hat Q_4^{}$ for two incoming electrons propagating along the chiral edges toward the superconducting beam splitter. Its expression reads
\begin{align}\label{Q3Q4_ee}
    &\langle \delta \hat Q_3^{} \delta \hat Q_4^{} \rangle_{ee}^{} (\tau) = -\text{Re} \sum_{\gamma,\delta \in \{e,h\}} q_\gamma^{} q_\delta^{} \Big[ \\
    &\phantom{\times} 2 \int d\varepsilon  |{\phi(\varepsilon)}|^2 (r^{\gamma e}_{31}(\varepsilon))^* t^{\gamma e}_{32}(\varepsilon) \text e^{\text i\varepsilon\tau} \notag \\
    &\times \int d\varepsilon' |{\phi(\varepsilon')}|^2 \left(r^{\delta e}_{42}(\varepsilon') \right)^*  t^{\delta e}_{41}(\varepsilon')\ \text e^{-\text i\varepsilon'\tau} \notag \\
    &+\int d\varepsilon |t^{\gamma e}_{32}(\varepsilon)|^2|\phi(\varepsilon)|^2 \int d\varepsilon' |r^{\delta e}_{42}(\varepsilon')|^2|\phi(\varepsilon')|^2 \notag \\
    &+\int d\varepsilon |t^{\delta e}_{41}(\varepsilon)|^2 |\phi(\varepsilon)|^2  \int d\varepsilon' |r^{\gamma e}_{31}(\varepsilon')|^2 |\phi(\varepsilon')|^2 \Big] \notag .
\end{align}
Equation~\eqref{Q3Q4_ee} consists of four contributions, corresponding to the possible charge configurations of the outgoing quasiparticles at the drain contacts 3 and 4,
\begin{align}\label{4terms_ee}
    \langle \delta \hat Q_3 \delta \hat Q_4 \rangle
    &= \langle \delta \hat Q_3 \delta \hat Q_4 \rangle_{ee}^{ee}
    + \langle \delta \hat Q_3 \delta \hat Q_4 \rangle_{ee}^{he} \notag \\
    &+ \langle \delta \hat Q_3 \delta \hat Q_4 \rangle_{ee}^{eh}
    + \langle \delta \hat Q_3 \delta \hat Q_4 \rangle_{ee}^{hh}.
\end{align}
The term $\langle \delta \hat Q_3 \delta \hat Q_4 \rangle_{ee}^{ee}$ corresponds to normal processes, while the three remaining terms involve at least one Andreev reflection and one Andreev transmission. For $\Delta = 0$, when only normal processes take place, only the first term $\langle \delta \hat Q_3 \delta \hat Q_4 \rangle_{ee}^{ee}$ contributes to the charge covariance, and we recover the result of the previous section for a purely electronic beam splitter. This quantity corresponds to the time-integrated current cross-correlations measured in electron quantum optics experiments. In this case, the charge covariance (which is then equal to $\langle \hat Q_3^{} \hat Q_4^{} \rangle - 1$) is always negative and has a maximum at $\tau \approx 0$. 
Furthermore, if the amplitudes are constant, $\langle \delta \hat Q_3 \delta \hat Q_4 \rangle_{ee}^{ee}$ is greater than or equal to $-2 |r_{31}^{ee}|^2 |t_{32}^{ee}|^2$, reaching 0 at its maximum. For $\Delta_\text{S}^{} >0$, a direct signature of local and cross-Andreev reflections can be identified in the HOM signal: 
the HOM peak orients downwards and changes from a maximum to a minimum. The reason for this change can be elucidated by starting from the unitarity of the scattering matrix extended to the hole sector. Indeed, the analog of the third line of equation \eqref{scatt_cond} for a BdG scattering matrix now becomes
\begin{equation}\label{scatt_cond_SC}
    (r_{31}^{ee})^* t_{32}^{ee} + r_{42}^{ee} (t_{41}^{ee})^* + (r_{31}^{he})^* t_{32}^{he} + r_{42}^{he} (t_{41}^{he})^* = 0.
\end{equation}
If Andreev processes are absent, we recover the third line in Eq.~\eqref{scatt_cond}. If the HOM peak is downward-orientated, in contrast, it means that the unitarity relations in Eq.~\eqref{scatt_cond} alone are not satisfied and that at least one of the terms $(r_{31}^{he})^*t_{32}^{he} + r_{42}^{he}(t_{41}^{he})^*$ in Eq.~\eqref{scatt_cond_SC} must be non-zero. This can only be the case if at least one Andreev amplitude is non-zero and thus can be used to detect their presence.

\begin{figure}
    \centering
    \includegraphics[width=\linewidth]{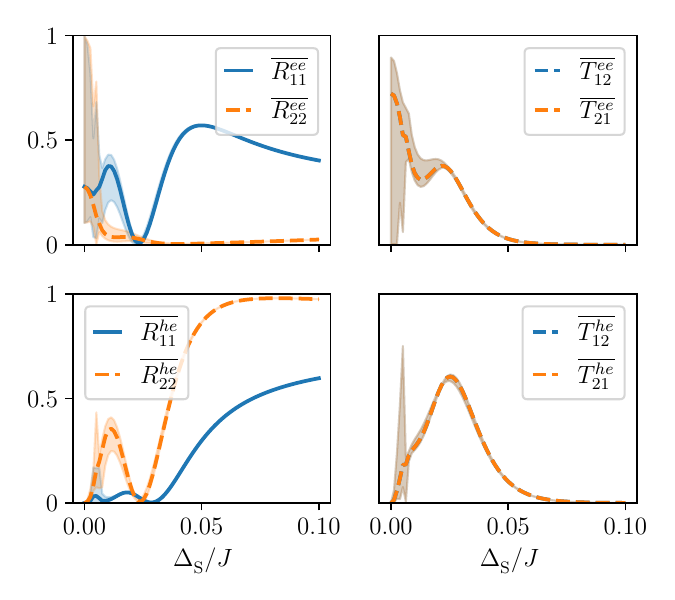}
    \caption{Scattering probabilities $|S_{ij}^{\mu e}|^2$ averaged over the energy window $[\varepsilon_0-3\sigma,\varepsilon_0+3\sigma]$ as a function of the superconducting gap $\Delta_\text{S}$. The shaded areas indicate the corresponding minimum-to-maximum variation within that energy window.}
    \label{fig:Scatt_Amplitudes_avg}
\end{figure}

\begin{figure}[t]
    \centering
    \includegraphics[width=\linewidth]{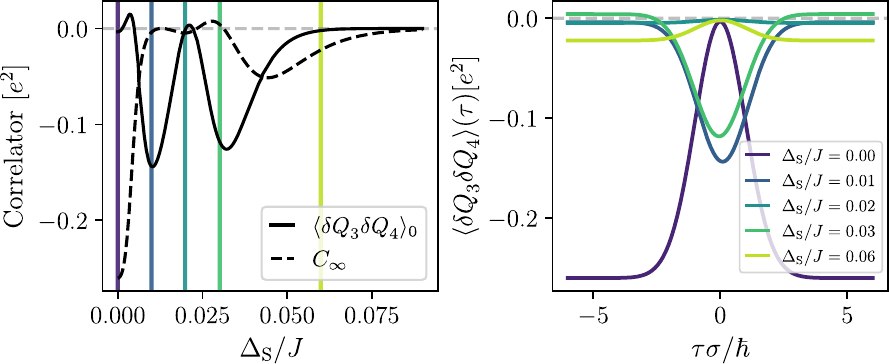}
    \caption{\textit{a)} Value of the charge covariance at its peak/dip for $\tau \approx 0$ (solid black line) and at $\tau = \pm \infty$ (black dashed line) versus the superconducting gap $\Delta_\text{S}^{}$. \textit{b)} Charge covariance versus the time delay for different values values of $\Delta_\text{S}^{}$. The colors of the lines correspond to those of the vertical lines in panel \textit{a}.}
    \label{fig:dQ3dQ4_SC}
\end{figure}

A simple analytical understanding can be gained from the observation that, for $\Delta_\text{S} \gtrsim 0.015J$, the scattering amplitudes vary only weakly across the energy window of the wavepacket, as shown in Fig.~\ref{fig:Scatt_Amplitudes_avg}, where we plot the scattering probabilities
\begin{equation}
    \overline{|S_{ij}^{\alpha \beta}|^2} = \frac{1}{6\sigma} \int_{\varepsilon_0^{}-3\sigma}^{\varepsilon_0^{} + 3\sigma} |S_{ij}^{\alpha \beta}(\varepsilon)|^2 d\varepsilon
\end{equation}
averaged over the interval $[\varepsilon_0-3\sigma,\varepsilon_0+3\sigma]$, together with the spread they retain within that window. In this regime, each contribution to Eq.~\eqref{4terms_ee} can be approximated by
\begin{align}
    & \langle \delta \hat Q_3^{} \delta \hat Q_4^{} \rangle^{\alpha \beta}_{ee}
    = -q_\alpha^{} q_\beta^{}
    \Bigg[ 
    |t_{32}^{\alpha e}|^2 |r_{42}^{\beta e}|^2
    + |t_{41}^{\alpha e}|^2 |r_{31}^{\beta e}|^2 \notag \\
    & +2 |r_{31}^{\alpha e}| |r_{42}^{\beta e}| |t_{32}^{\alpha e}| |t_{41}^{\beta e}| \times \\
    & \times \text{Re}
    \int d\varepsilon\, |\phi(\varepsilon)|^2
    e^{i(\varphi_3^\alpha(\varepsilon)+\varepsilon\tau)} 
    \int d\varepsilon'\, |\phi(\varepsilon')|^2
    e^{i(\varphi_4^\beta(\varepsilon')-\varepsilon'\tau)}
    \Bigg] \notag,
\end{align}
with
\begin{equation}
    \begin{split}
    \varphi_3^\alpha(\varepsilon) &=
    \arg\!\left[t_{32}^{\alpha e}(\varepsilon)\right]
    - \arg\!\left[r_{31}^{\alpha e}(\varepsilon)\right],\\
    \varphi_4^\beta(\varepsilon) &=
    \arg\!\left[t_{41}^{\beta e}(\varepsilon)\right]
    - \arg\!\left[r_{42}^{\beta e}(\varepsilon)\right].
    \end{split}
\end{equation}
In this approximation, the first two terms are independent of $\tau$ and therefore fix the asymptotic value
\begin{equation}
\begin{split}
    C_\infty &\equiv  \sum_{\alpha,\beta}^{}\langle \delta \hat Q_3 \delta \hat Q_4 \rangle_{ee}^{\alpha \beta}(\tau \to \pm \infty),\\
    &= - \sum_{\alpha, \beta} q_\alpha^{} q_\beta \left[|t_{32}^{\alpha e}|^2 |r_{42}^{\beta e}|^2
    + |t_{41}^{\alpha e}|^2 |r_{31}^{\beta e}|^2\right]
\end{split}
\end{equation}
of all contributions with $\alpha,\beta \in \{e,h\}$ for $|\tau| \to \infty$. The last term contains the HOM interference and is proportional to the amplitudes. This term determines whether the corresponding contribution appears as a dip or a peak around $\tau=0$. The total charge covariance is shown in Fig.~\ref{fig:dQ3dQ4_SC} for several values of the superconducting gap $\Delta_\text{S}$, chosen to illustrate the different regimes of the signal. In the normal limit ($\Delta_\text{S}=0$, purple line), the result reduces to that of a purely electronic HOM interferometer and remains negative for all time delays $\tau$.

As $\Delta_\text{S}$ increases, the covariance progressively departs from this behavior. In panel~\textit{a}, the asymptotic value (black dashed line) and the asymptotic value (black solid line) approach each other and eventually cross, leading to an inversion of the HOM signal. This corresponds to a downward-orientated feature, as illustrated for $\Delta_\text{S}=0.01J$ (dark blue line in panel~\textit{b}). Upon further increasing the gap, additional inversions occur: first around $\Delta_\text{S}\approx 0.02J$ (turquoise), and again in the range $0.02J<\Delta_\text{S}<0.04J$ (shown for $\Delta_\text{S}=0.03J$, green). For larger gaps ($\Delta_\text{S}>0.04J$), the signal stabilizes in an upward orientation, as seen for $\Delta_\text{S}=0.06J$ (lime green). At the same time, the overall amplitude decreases and eventually tends to zero, consistent with the exponential suppression of both normal and Andreev transmission processes.

As motivated above, we focus on a regime where the superconducting region length is comparable to the coherence length, namely $L_{\text{SC}} \simeq 2 \xi_0$. In the simulations presented in Fig.~\ref{fig:dQ3dQ4_SC}, we fix $L_{\text{SC}} = 104a$, which corresponds to targeting a gap of $\Delta_{\text{S}} \simeq 0.03 J$. This choice places the system in a regime where cross-Andreev reflection is expected to be significant, as evidenced by the downward orientation of the covariance signal.

The change of orientation of the HOM signal is thus a strong signature of crossed Andreev processes in the superconducting beam splitter. We note that an initial state containing a particle in one contact and a hole in the other might give a similar signature, even in the absence of Andreev processes. However, in this case, the injected particle would have to comprise energy eigenstates within a window of $k_B T$ around the energy of the (thermal) hole, a process which is unlikely at small temperatures $k_B T \ll \varepsilon_0$. Coulomb interactions along the edge can cause charge fractionalization and plasmon-mediated wavepacket spreading \cite{wahlInteractionsChargeFractionalization2014}. This results in a reduced visibility of the HOM signal but is not expected to change the qualitative sign inversion driven by Andreev processes.

\section{Conclusion}

We have investigated a superconducting analogue of an electronic Hong-Ou-Mandel (HOM) interferometer, where the beam splitter is implemented as a narrow superconducting wire. We have numerically calculated the scattering matrix of the system and used it to calculate the interference signatures for an initial state consisting of two particles with a time delay. In the normal conducting case, the cross correlation of the outgoing currents displays the well-known HOM peak as a result of fermionic anti-bunching. We showed that Andreev processes in the case of a superconducting beam splitter strongly modify this signature and can lead to an inversion of the HOM peak. We have shown that such an inversion could be used to detect and characterize Andreev processes in superconductor-quantum Hall interfaces.

\begin{acknowledgments}
TM and TLS acknowledge financial support from the National Research Fund Luxembourg under grant INTER/QUANTERA21/16447820/MAGMA and PRIDE/23/18691647/QUANCOM. MK acknowledges funding from the Research Council of Finland under grant no 338872 (NAI-CoG) and QuantERA II Program that has received funding from the European Union’s Horizon 2020 Research and Innovation Programme under Grant Agreement Nos 731473 and 101017733. The authors acknowledge fruitful discussions with Bernd Braunecker, Alessandro De Martino, Christophe Mora, and Patrik Recher.
\end{acknowledgments}

\bibliography{references}

\end{document}